\documentclass{article}

\usepackage{PRIMEarxiv}

\usepackage[utf8]{inputenc} 
\usepackage[T1]{fontenc}    
\usepackage{hyperref}       
\usepackage{url}            
\usepackage{booktabs}       
\usepackage{amsfonts}       
\usepackage{nicefrac}       
\usepackage{microtype}      
\usepackage{lipsum}
\usepackage{tikz}
\usetikzlibrary{arrows.meta, positioning}
\usepackage{hyperref}
\usepackage{algorithm}
\usepackage{amsmath}
\usepackage{multirow}
\usepackage{algorithmic}
\usepackage{graphicx}
\usepackage{textcomp}
\usepackage{xcolor}
\usepackage{caption}
\usepackage{subfig}

\usepackage{fancyhdr}       
\usepackage{graphicx}       
\graphicspath{{media/}}     

\title{HICM: An approach towards Harmonizing Indian Census Migration data and its applications
}

\author{
  Niveditta Batra \\
  Computer Science and Engineering \\
  Indian Institute of Technology \\
  Jodhpur, India\\
  \texttt{niveditta.2@iitj.ac.in} \\
   \And
  Chiranjoy Chattopadhyay \\
  School of Computing and Data Sciences \\
  FLAME University\\
    Pune, India \\
  \texttt{chiranjoy.chattopadhyay@flame.edu.in} \\
  \And
  Mayurakshi Chaudhuri \\
  School of Liberal Education \\
  FLAME University\\
    Pune, India \\
  \texttt{mayurakshi.chaudhuri@flame.edu.in} \\
}

\begin{document}
\maketitle

\begin{abstract}
Reliable analysis of migration is critically dependent on the quality and consistency of the underlying data. Indian migration data, primarily derived from decennial census records, are affected by systematic gaps arising from uneven coverage and measurement inconsistencies across states and time. This paper presents a data-centric framework, HICM, for harmonizing Indian census migration data recorded under the Indian census and correcting prominent sources of bias prior to downstream analyses. We explicitly identify two types of bias across three decades of migration data: measurement bias and representativeness bias. We propose to address these gaps through principled pre-processing, mitigation, and validation strategies grounded in statistical diagnostics. An empirical evaluation of harmonized Indian interstate migration data reveals that bias-aware data correction substantially improves the consistency in the structure of the data and enhances the reliability of subsequent temporal analysis results. By improving data quality through reproducible data imputation and smoothing, this work advances migration analytics and provides a robust foundation for policy-relevant longitudinal network analysis of Indian internal migration.
\end{abstract}


\section{\textbf{Introduction}}

Systematic biases and inconsistencies in Indian Census data have been widely acknowledged and examined in prior literature. Studies have documented multiple sources of census error, including coverage errors \cite{kumar2023census} \cite{kumar2020census}, definitional ambiguities \cite{sircar2017census}, age misreporting \cite{yadav2020quality} \cite{chattopadhyay2024age}, representativeness bias \cite{singh2025counting} \cite{singh2025caste} \cite{ranjan2025caste}, and inconsistencies arising from changes in administrative boundaries and data collection practices over time. Correspondingly, a range of statistical and mathematical imputation strategies, such as proportional redistribution, marginal adjustment, iterative proportional fitting, and temporal interpolation, have been proposed and applied to demographic and socioeconomic census variables to mitigate missingness and reporting distortions \cite{starr2021measuring} \cite{datta2024spatial}.

In contrast, data missingness intrinsic to internal migration data recorded in the Indian Census remains comparatively understudied, despite its central role in understanding population displacement, urbanization, and regional inequality \cite{datta2013indian} \cite{bhagat2025nature}. The Indian Census offers a uniquely rich longitudinal resource, reporting migration stocks at decennial intervals over more than three decades. This periodic and consistent data collection enables longitudinal trend analysis and the study of structural shifts in migration patterns \cite{batra2024computational} \cite{islam2024enquiry}. However, the analytical potential of this multi-decadal migration data is often undermined by biases caused by structural inconsistencies and missingness that are not explicitly addressed by existing studies \cite{garha2019indian}, \cite{srivastava2011internal}.

Migration-specific census data present a distinct set of gaps that differ fundamentally from those encountered in standard population counts. Measurement gaps arise from under-reporting migration attributes across census rounds, unclassifiable places of last residence, and missing duration-of-stay variables \cite{srivastava2013impact}. Representation bias manifests through the underrepresentation of temporary, circular, or informal mobility, often compounded by gendered reporting practices, and the underrepresentation of politically challenged states \cite{kundu2019trends}. These biases distort both the magnitude and structure of observed migration patterns, thereby affecting statistical, temporal, and network-based analyses. This paper, to the best of our knowledge, is the first to offer the following contributions: 

\begin{enumerate}
    \item Bias profiling through existing gaps in the interstate migration data collected by the Indian Census.
    \item Mathematical strategies to correct identified gaps, offering a harmonization of multi-decadal Census migration datasets for downstream analyses
\end{enumerate}

We adopt a systematic data-centric perspective, emphasizing the identification, formalization, and correction of bias at the data level. All data tables for the years 1991, 2001, and 2011 are publicly provided by the Indian census. In this paper, we focus specifically on the D-02 series \footnote{Accessible at \url{https://censusindia.gov.in/census.website/data/census-tables}} of Census migration tables, which reports migration stocks by source state, destination state, rural–urban classification and duration of residence. This series of data and its proposed harmonization strategy are especially valuable for network-oriented migration studies, as it supports the construction of directed, weighted migration networks and enables temporally grounded analyses across census decades. Our paper focuses specifically on interstate migration in India in all origin-destination state pairs. We have excluded movements to other countries and inter-district movements in this paper.

The remainder of the paper is structured as follows: Section \ref{Problem} discusses the problem statement, providing three categories of biases and inconsistencies with respect to the Indian census migration data. Section \ref{strategy} discusses a detailed harmonization strategy for each of these biases. Next, in section \ref{result}, we discuss the validation of these strategies. This is followed by a discussion in section \ref{discussion} and section \ref{conclusion} concludes the paper.

\section{\textbf{Problem Statement}} \label{Problem}
We identify the following gaps specific to the Indian migration data recorded under the Census of India for three different decades: 1991, 2001, and 2011. The data prior to 1991 is not available in digital form, and the census after 2011 has not been conducted yet, hence not included in our paper. 

\subsection{Unclassifiable/ Unknown Measurements}
We identify two sources of measurement gaps in the datasets: (i) Migrants with unclassifiable last residence, (ii) Migrants with duration of stay unknown. The following subsections discuss these two categories in detail. 

\begin{table}[t]
\centering
\caption{Percentage share of migrants: (A) Intrastate (B) Interstate (C) International and (D) Unclassifiable, across three decades with respect to total migrants}
\label{tab:unclassifiable_migrants}
\begin{tabular}{l|c|c|c|c}
\textbf{Year} & \textbf{Intrastate}&\textbf{Interstate}&\textbf{International}&\textbf{Unclassifiable} \\ \hline
2011 &86.806\% &11.905\%&1.204\% &0.083 \%  \\
2001 &85.273\% &13.087\%&1.639\% &0.0001 \%  \\
1991 & 85.819\%&11.498\%& 2.553\%&0.128 \% \\
\end{tabular}
\end{table}

\subsubsection{Unclassifiable migrants}The D-02 Indian census provides the migrant stocks residing in states and union territories, i.e., in the ``Places of enumeration'' by dividing the migrant stocks into four categories on the basis of last residence: (A) Within the state of enumeration but outside the place of enumeration (further divided as 1. Elsewhere in the district of enumeration, 2. In other districts of the state of enumeration), (B) States in India beyond the state of enumeration, (C) Last residence outside India, and (D) Unclassifiable. The individual share of each of these categories in each decade data is summarized in Table \ref{tab:unclassifiable_migrants}. For the ease of reading, we call these categories (A) Intrastate, (B) Interstate, (C) International, and (D) Unclassifiable migrants. 

As per the census data, “Unclassifiable migrants" record the stocks of migrants in every place of enumeration, whose ``last residence'' is unknown. These unclassifiable migrants are present across all three decades of the census, and these migrants hold a total share of $<$ 1\% of the total migrants in India for each decade. This may lead to under-reporting of these migrants or omission in certain analyses (e.g., presence of unclassifiable migrants may cause under-counts in out-migrant strength, thereby biasing the net migration balance toward places with a high in-migrant strength). 

\begin{table}[b]
\centering
\caption{Decade-wise percentage share held by Indian migrants as per durations of stay, D, in years}
\label{tab:duration_not_stated}
\begin{tabular}{l|c|c|c|c}
\textbf{Year} & \textbf{D $<1$}& \textbf{$1\leq$ D $\leq20$} & \textbf{$20<$D}& \textbf{D not stated} \\ \hline
2011 & 3.858\% &48.058\%&32.124 \% & 15.960 \% \\
2001 & 2.824\%& 50.516\% &32.139\%&14.521 \% \\
1991 & 3.047\%& 55.851\% &32.677\% &8.425 \% \\
\end{tabular}
\end{table}
\subsubsection{Unknown duration of stay} The Indian census categorizes the duration of stay in the place of enumeration into five distinct categories: “Less than 1 year", “1-4 years", “5-9 years", “10-19 years", and “20+ years". However, there is one additional category, “Duration not stated"/“Period not stated", and such migrants with duration not stated are present across all three datasets. Each source-destination state pair holds its individual share of such migrants the duration of stay is not known. Table \ref{tab:duration_not_stated} showcases the decade-wise share of migrants as per duration bins, containing the shortest duration (duration of stay less than 1 year), cumulative intermediate bins (duration of stay 1 to 20 years), duration more than 20 years, and duration not stated. Across each decade, the share of unknown duration of stay migrants account for $\approx 8-16 \%$ of the total migrants in India.  This is a non-trivial missingness that can materially affect certain analyses. 

With this, we have two missingness profiles with missingness percentage, M, across all three datasets: (i) Migrants with unclassifiable source (where M$<$ 5\%), (ii) Migrants with duration of stay unknown (where 5\% $\leq$ M $<$ 20\%).

\subsection{Coverage inconsistency}
The 2001 and 2011 census reports the migrants residing in 35 states and union territories as places of enumeration. However, in the case of the 1991 census dataset, there has been an evident inconsistency in terms of the coverage of all states and union territories. The dataset records the migrant population in India, excluding migrants residing in Jammu and Kashmir (J\&K) due to the political instability in 1991 \cite{kumar2024numbers}. On the contrary, the migrants outflow data of J\&K is available for all three decades. As per the 2011 and 2001 census data, J\&K holds 0.62\% and 0.57\% of the total migrant inflow population, respectively. Considering India's vast population, excluding these percentages, or not having an entire state's in-migrant population in longitudinal studies, leads to representation bias in analyses against the migrants residing in an unrecorded state. This absence in the 1991 dataset also leads to inconsistencies in the source-destination pairs for Indian internal migration for studies, such as network-specific analysis.

\subsection{Nomenclature/ Indexing inconsistency} Several of the inconsistencies related to naming or numerical indexing of the data identified within the datasets are summarized below: 
\subsubsection{Indexing Mismatch} A numerically ordered index is assigned to the states in the 1991 dataset, which is done alphabetically, first for all the states, and second, for all the union territories, all in one continuum. For example, the state Andhra Pradesh is assigned a numerical label ``(1)'' and West Bengal as ``(25)'', followed by the union territory Andaman and Nicobar as ``(26)'', and so on. This indexing does not align with the indexing used in the other two decades' datasets, which are done geographically, i.e., the geographically northern state, Jammu and Kashmir is assigned as the label ``(1)'', and not as per its alphabetical order. This inconsistency, if gone unchecked, can potentially lead to propagation of silent computational errors, especially in automated analyses pipelines. Inconsistent numeric labels require additional preprocessing logic and increasing methodological complexity. 

\subsubsection{Total column missing} The 1991 dataset lacks an aggregated total value column per record, indicating incomplete summary information on migrant information. The 2001 and 2011 Census datasets include explicit total columns, and the lack of corresponding summary information in 1991 introduces structural asymmetry that complicates direct cross-decade comparisons. Aggregated totals are essential as they also serve as hard constraints during redistribution or imputation. Their absence weakens the ability to enforce sum-preserving corrections and increases reliance on inferred totals. 

\subsubsection{Name Mismatch} Variant state names, such as “Orissa"/“Odisha", “A\&N Islands"/“Andaman and Nicobar Islands", etc., have been observed across all three datasets. In a study where the state names are considered as the primary identifiers across all the datasets for temporal comparisons, and if these identifiers do not correspond consistently across datasets, a direct decade-to-decade comparisons may attribute changes in migration volumes, or hold an additional risk of undocumented assumptions.

\begin{figure}[t]
\centering
\resizebox{!}{0.1\textwidth}{
\begin{tikzpicture}[
    timeline/.style={line width=6pt, draw=primary!80, -{Stealth[length=5mm]}},
    milestone/.style={
        rectangle,
        rounded corners=2pt,
        minimum width=6cm,
        minimum height=0.8cm,
        align=center,
        font= \sffamily,
        text width=5cm,
        anchor=center
    },
    yearbox/.style={
        rectangle,
        rounded corners=2pt,
        fill=gray!10,
        draw=gray!30,
        minimum width=11cm,
        minimum height=0.8cm,
        align=center,
        font= \sffamily,
        text width=7cm
    },
    primary/.style={fill=blue!20, draw=blue!50},
    secondary/.style={fill=green!20, draw=green!50},
    tertiary/.style={fill=orange!20, draw=orange!50},
    quaternary/.style={fill=purple!20, draw=purple!50},
    connector/.style={
        line width=0.8pt,
        draw=gray!40,
        -{Stealth[length=3mm]}
    }
]

\definecolor{primary}{RGB}{70,130,180}
\definecolor{secondary}{RGB}{60,179,113}
\definecolor{tertiary}{RGB}{255,165,0}
\definecolor{quaternary}{RGB}{147,112,219}

\newcommand{\milestone}[5]{%
  \node[milestone,#3,align=center] at (#1, #2) {#4};
}
\newcommand{\technique}[5]{%
  \node[yearbox,#3,align=center] at (#1, #2) {#5};
}

\milestone{0.1}{6.5}{quaternary}{Mitigating Nomenclature/ Label/ Aggregation inconsistency}{}
\draw[-{Stealth[length=3mm]}] (3.1,6.5) -- (4.4,6.5);
\milestone{7.4}{6.5}{quaternary}{Mitigating coverage inconsistency}{}
\milestone{7.4}{4.5}{quaternary}{Redistributing unclassifiable migrants to ``Last" residence}{}
\milestone{0.1}{4.5}{quaternary}{Redistributing ``unknown duration of stay" migrants}{}

\draw[-{Stealth[length=3mm]}]  (4.4,4.5) -- (3.1,4.5);

\draw[-{Stealth[length=3mm]}] (7,6) -- (7,5);


\end{tikzpicture}
}
\caption{Flow diagram summarizing the process of applying harmonization strategies on various bias profiles.}
\label{fig:methodology}
\end{figure}
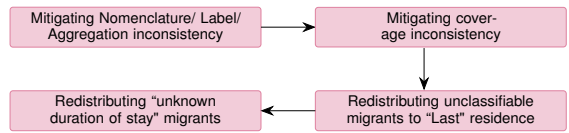

\section{\textbf{Harmonization Strategy}} \label{strategy}
Fig. \ref{fig:methodology} summarizes the overall flow of the solution. First, we perform fixing of nomenclature and sequential indexing inconsistencies across all the datasets, as discussed in the subsection \ref{nomenclature_Edits}. Second, to ensure inter-decade consistency, we perform temporal smoothing, which eliminates representation gaps specific to the 1991 migration dataset as discussed in subsection \ref{Enumeration_Imputation}. Third, we apply imputation to datasets across all three decades, ensuring that complete and usable datasets are available for each decade to support decade-specific studies. This is discussed in steps 1 and 2 of the subsection \ref{combined_imputation}.

\subsection{Mitigation of Nomenclature/ Label/ Aggregation inconsistencies} \label{nomenclature_Edits} The following corrections address nomenclature and indexing inconsistencies, thereby enabling consistent alignment with standardized identifiers across all census decades:
\begin{enumerate}
    \item Following a standardized indexing of states in the 1991 dataset, as per the 2001 and 2011 datasets. This harmonization eliminates ambiguities arising from inconsistent numerical labeling, ensures correct alignment of places of enumeration. 
    \item Listing a “Total" migrant count column against each duration of stay in the 1991 dataset. This step ensures dimensional consistency with the 2001 and 2011 datasets, enables internal consistency checks between disaggregated and aggregated values
    
    \item Normalization of state nomenclature to the latest identifiers. Replacing “Orissa" with “Odisha", ``Delhi'' with ``NCT of Delhi", “A\&N islands" with “Andaman and Nicobar Islands", “Pondicherry" with “Puducherry", and “Uttaranchal" with “Uttarakhand" in the 2001 and 1991 datasets was performed to ensure the nomenclature consistency. 
\end{enumerate}

\begin{algorithm}[t]
\caption{Temporal Ratio Smoothing for Backward Migration Flow Estimation}
\label{alg:temporal_ratio_smoothing}
\begin{algorithmic}[1]

\REQUIRE Migration flow matrices 
$\mathbf{F}^{1991}, \mathbf{F}^{2001}, \mathbf{F}^{2011}$,
where $f^{t}_{i,j}$ denotes migration from source $i$ to destination $j$ in decade $t$ obtained from the dataset ${F}^{t}$
\REQUIRE Missing place of enumeration, $x$, in 1991 
\ENSURE Estimated migration flows 
$f^{1991}_{i, x}$ for $x$ and all $i \in P$ 

\FOR{each place of last residence $i$}

    \STATE compute 1991-2001 and 2001-2011 out-migration ratio:
    \[
    R_i^{(1991,2001)} =
    \frac{\sum_{j} f^{1991}_{i,j}}
         {\sum_{j} f^{2001}_{i, j}}
    \]
    
    \[ R_i^{(2001,2011)} =  \frac{\sum_{j} f^{2001}_{i, j}}     {\sum_{j} f^{2011}_{i, j}}\]
    
    \STATE compute smoothed backward temporal ratio:
    \[
    \tilde{R}_i =
    \sqrt{
    R_i^{(1991,2001)} \cdot R_i^{(2001,2011)}
    }
    \]
\ENDFOR

\FOR{each place of last residence $i$}
        \STATE estimate missing 1991 flow:
        \[
        f^{(1991)}_{i, x}
        =
        f^{(2001)}_{i, x} \cdot \tilde{R}_i
        \]
\ENDFOR

\RETURN updated ${F}^{1991}$

\end{algorithmic}
\end{algorithm}

\subsection{Mitigation of Coverage Inconsistency} \label{Enumeration_Imputation}
Inconsistencies related to an unreported place of enumeration in a particular dataset can be treated by calculating the respective migrant flows from all the places of last residence (represented by a set \textit{P}) to the missing place of enumeration, \textit{x}. This is based on the assumption that migration occurs within stable systems of origin-destination linkages, and these linkages change gradually, not abruptly \cite{tumbe2012migration}, \cite{das2015nativity}, \cite{sahai2023social}. Hence, the ratios of migration over time are more stable than raw flows. For this, we perform a backward projection of counts by using a temporal transfer ratio by a two-step transfer ratio smoothing to calculate the migrant counts. 

We consider three datasets as three matrices, $F^{1991}$, $F^{2001}$ and $F^{2011}$. Let $f_{i,j}^t$ is the element in the dataset $F^{t}$, and it represents the migrant count from a place of last residence \textit{i}, to the place of enumeration \textit{j}, where \textit{i}$\in P$ and \textit{t}$\in\{{1991, 2001, 2011\}}$. In the 1991 dataset, there are 31 states and union territories as the places of enumeration (\textit{j}), hence \textit{j}$\in\{{1, \dots, 31\}}$. For a two-step temporal smoothing using the available datasets, we follow Algorithm \ref{alg:temporal_ratio_smoothing}, which provides the place of last residence-wise migrant inflow counts to the missing place of enumeration. Here, we measure out-migration temporal ratio, $\tilde{R}_i$, for each last place of residence, \textit{i} (where \textit{i}$\in$ \textit{P}). The value $\tilde{R}_i$ is the geometric mean of two cross-decades out-migration ratios, $R_{i}^{1991,2001}$ and $R_{i}^{2001,2011}$. All these ratios are specific to each last place of residence, from which the migrant count to the missing state has to be calculated. 

\begin{figure}[t] 
    \centering
  
  \subfloat[Gujarat\label{Gujarat}]{%
        \includegraphics[width=0.45\linewidth]{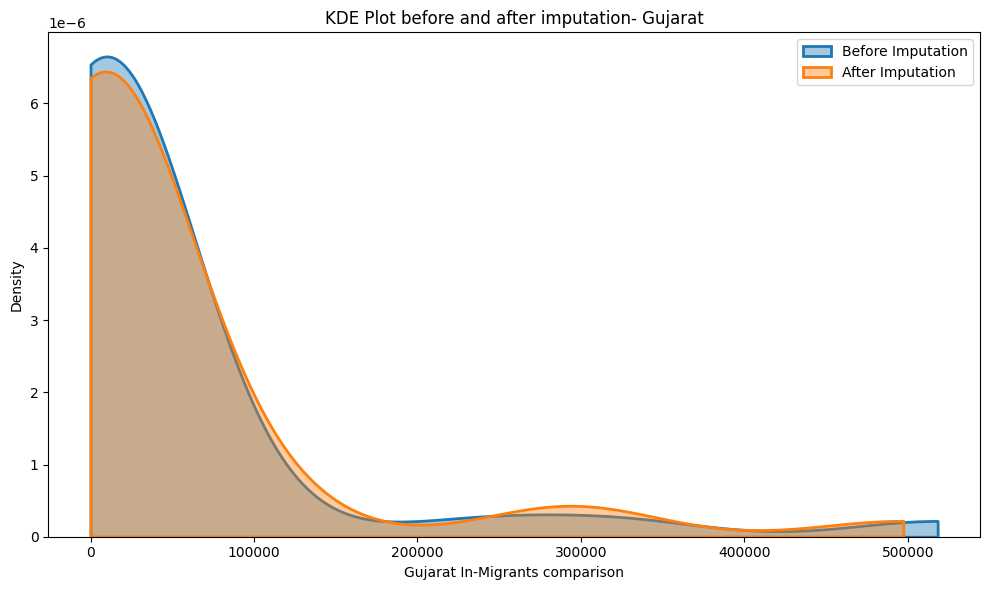}}
         \subfloat[Delhi\label{Delhi}]{%
        \includegraphics[width=0.45\linewidth]{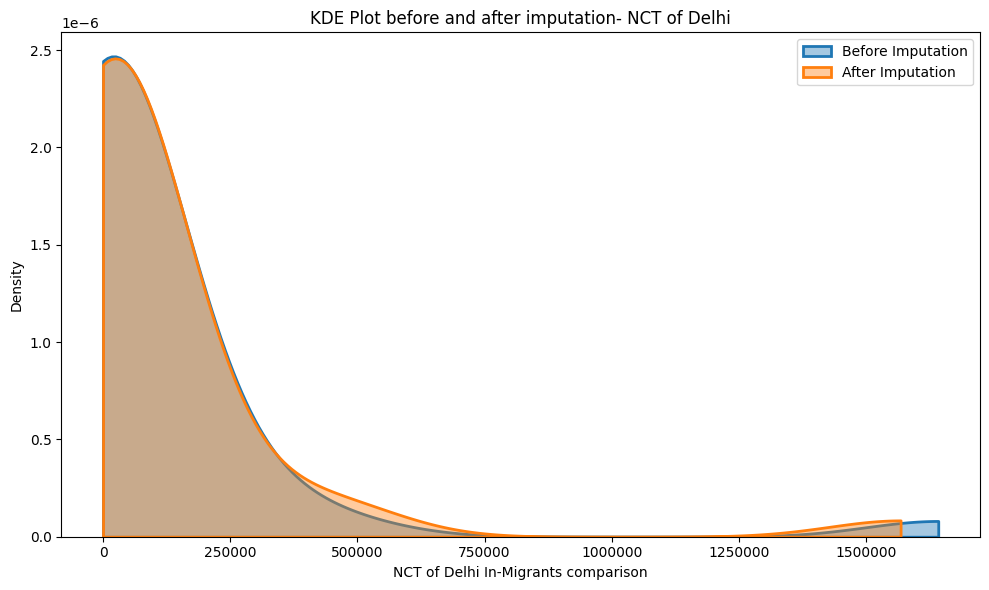}}
  \caption{Comparison of 1991 migrant inflow population: Ground truth vs Imputed data}
  \label{in flow violin plot} 
\end{figure}

We also attempt to estimate the in-migrant data distribution of the pre-existing places of enumeration in the 1991 dataset to ensure the quality of the estimated results using the proposed algorithm. To do so, we randomly chose two states that have been among the top five recipients of migrants in the latest migration dataset (2011). A comparison of the imputed share of these states' inflow in 1991 vs the actual shares recorded in 1991 is presented in Fig. \ref{Gujarat} and \ref{Delhi}. The imputed results show overlapping kernel density estimate plots, indicating a smooth replication of the actual 1991 dataset.

\subsection{Mitigation of Unclassifiable/ Unknown Measurements} 
\label{combined_imputation}

\subsubsection{Redistributing unclassifiable migrants to places of last residence}
Given a very small missing share ($<$ 1\% of the total migrants, refer Table \ref{tab:unclassifiable_migrants}), we redistribute these unclassifiable migrants to each of the last places of residence. Let $M_{ij}$ denotes the observed number of migrants from the last state of residence, $i$ to the place of enumeration, $j$,. For each place of enumeration $j$, let $U_j$ represent the number of migrants in $j$ whose place of last residence is reported as unknown. 

Since this missingness percentage in each dataset is very small, for each dataset, we perform a place of last residence-wise proportional redistribution of the unclassifiable-source migrants. This is calculated through the ratio of migrants coming from ``last place" to the total migrants in a place of enumeration. While preserving the relative migrant inflow in each place of enumeration,  we allocate the unclassifiable migrants to the known places of last residence in proportion to their observed migration flows into the corresponding place of enumeration. The redistributed migration count $\hat{M}_{ij}$ is calculated as:
\begin{equation}
\hat{M}_{ij} = M_{ij} + U_j \cdot \frac{M_{ij}}{\sum\limits_{i} M_{ij}},
\quad \text{for } \sum\limits_{i} M_{ij} > 0
\end{equation}

In each destination state, $d$, the value $\sum\limits_{i} M_{ij}$ will always be non-zero, as each destination state has reported migrant inflow in each decade. Hence, it is not required to handle the zero-denominator error while applying this distribution. This formulation would also ensure preserving the overall total migrant count and place of enumeration-wise total migrants before and after imputation. 
\subsubsection{Redistributing ``unknown duration of stay" migrants} If the migrants reported under 'duration not available' ($\approx 8-16 \%$) are redistributed across duration categories under the same assumption of duration-wise proportional redistribution, then it would reinforce the existing distribution, masking structural undercounts. We cannot perform a direct temporal ratio transfer across decades either, as this gap persists across all three decades' datasets. Hence, we encode the sociological assumption that missingness is largely higher among recent migrants, as reported by multiple studies \cite{coffey2015short} \cite{korra2012short} \cite{gupta2020study} \cite{borhade2025internal} \cite{bhagat2025nature}. 



Let $b \in \{1,2,3,4,5\}$ are the indices of given five migration duration bins 
($<1$ year, $1$--$4$ years, $5$--$9$ years, $10$--$19$ years, $20+$ years). We define a fixed weight vector ${w} = (w_1, w_2, w_3, w_4, w_5)$ such that it follows an exponential decay, $\tilde{w}_b = e^{(-\lambda (b - 1))})$, and satisfies Eq. \eqref{sum_of_weights}:
\begin{equation}
\sum_{b=1}^{5} w_b = 1,
\qquad
w_1 \ge w_2 \ge w_3 \ge w_4 \ge w_5
\label{sum_of_weights}
\end{equation}

Ensuring that ($w_1 + w_2 + w_3> 80\%$), taking $\lambda=0.4$, and keeping the long-duration tail ($w_5$) as non-zero, we can consider 
\begin{equation}
    w \approx (0.35,0.30,0.20,0.10,0.05)
\end{equation}

For a particular dataset, let $N_{j,b}$ be the reported number of migrants in the place of enumeration $j$ and duration bin $b$, and $\widehat{N}_{j,b}$ is the recalculated number of migrants for the same $j$ and $b$. Here, $j \in \{1,\dots,35\}$ for the 2001 and 2011 dataset, and $j \in \{1,\dots,32\}$ for 1991 dataset. The migrants reported under ``duration not stated", denoted by $U_{j}$, are redistributed to calculate $\widehat{N}_{j,b}$ as:
\begin{equation}
\widehat{N}_{j,b}
=
N_{j,b}
+
w_b \cdot U_{j},
\qquad \forall b \in \{1,\dots,5\}.
\end{equation}
This formulation encodes the assumption that recent migrants are more likely to have missing duration information while preserving the total migrant count,
i.e.,
\begin{equation}
\sum_{b=1}^{5}
\left(
\widehat{N}_{j,b} - N_{j,b}
\right)
=
U_{j}.
\end{equation}

These step-wise strategies, after being applied to the three available datasets, provide a harmonized version of Indian internal migration data while preserving the overall data constraints, such as migrant aggregates and data distributions in each of the datasets.

\section{\textbf{Experimental Results}}
\label{result}

After harmonizing the Indian interstate migration data, we obtained three imputed datasets, corresponding to the three decades: 1991, 2001, and 2011. This section compares the data distribution of the imputed datasets with the original datasets. A comparison is made between the distribution of the missing state enumeration in the 1991 dataset and the existing data distributions of 2001 and 2011. These cross-decade comparisons are presented through violin plots, as they simultaneously convey the full distributional shape and key summary statistics of all three decades. Additionally, violin plots are also used to compare the overall inflow and outflow populations before and after imputation. 




\subsection{After imputing in-migrant counts to the missing place of enumeration in the 1991 dataset}

\begin{figure}[t]
\centering
\resizebox{!}{0.3\textwidth}{
 \includegraphics[width=0.37\linewidth]{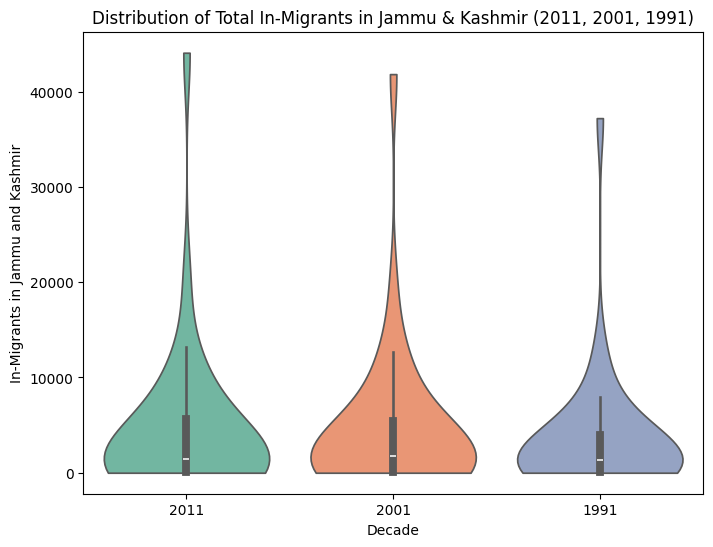}
}
\caption{Violin plot for decade-wise comparison of migrant inflow after imputing missing states data in the 1991 dataset}
\label{jammu violin plot}
\end{figure}

Algorithm \ref{alg:temporal_ratio_smoothing} provides a technique to estimate the in-migrant counts for a missing place of enumeration in the 1991 dataset. Using this, the missing Jammu and Kashmir inflow data are imputed and visualized in Fig. \ref{jammu violin plot}. The violin plots compare the data distribution across all three decades for Jammu and Kashmir. The 2011 dataset reports 0.62\% share of in-migrants held by Jammu and Kashmir, while the 2001 dataset reports 0.57\% in the same. The estimated counts for the 1991 dataset provide 0.399\% share, exhibiting no abrupt discontinuities or spikes, thereby indicating temporal consistency in the reconstructed in-migration patterns.



\begin{table}[b]
\centering
\caption{Comparison of Indian migrant counts with classifiable place of enumeration in the 2011 dataset before and after redistribution of ``Unclassifiable'' migrants. All values are reported in millions of migrants ($\times 10^{6}$).}
\label{tab:temporal_smoothness_step1}
\begin{tabular}{l|cc|cc}
\multirow{2}{*}{\textbf{Statistic}} 
&\multicolumn{2}{c|}{\textbf{In-migrant Distribution}} 
&\multicolumn{2}{c}{\textbf{Out-migrant Distribution}} \\ 
& \textbf{Before} & \textbf{After} & \textbf{Before} & \textbf{After} \\ \hline
Mean      & 1.550 & 1.553 & 1.550  & 1.553 \\
Std. Dev. & 2.007 & 2.011 & 2.411  & 2.416 \\
Min       & 0.006 & 0.006 & 0.016 & 0.016 \\
Q1        & 0.130 & 0.130 & 0.073 & 0.073 \\
Q2 (Median)& 0.654 & 0.654 & 0.694 & 0.694 \\
Q3        & 2.435 & 2.435 & 2.008 & 2.009 \\
Max       & 9.087 & 9.094 & 12.320 & 12.346 \\
\end{tabular}
\end{table}

\subsection{After redistributing unclassifiable and unknown duration of stay migrants}
In each dataset, the unclassifiable migrants present in each place of enumeration are redistributed across all places of last residence. This is implemented while preserving the total migrant population of each decade. We perform the redistribution across the three categories ((A) Intrastate, (B) Interstate, and (C) International) of places of last residence. Once the unclassifiable migrants are redistributed, the migrant population with ``Duration not stated" is redistributed across all the duration bins as well.

\begin{figure}[t] 
    \centering
  \subfloat[Inflow comparison\label{1a}]{%
       \includegraphics[width=0.37\linewidth]{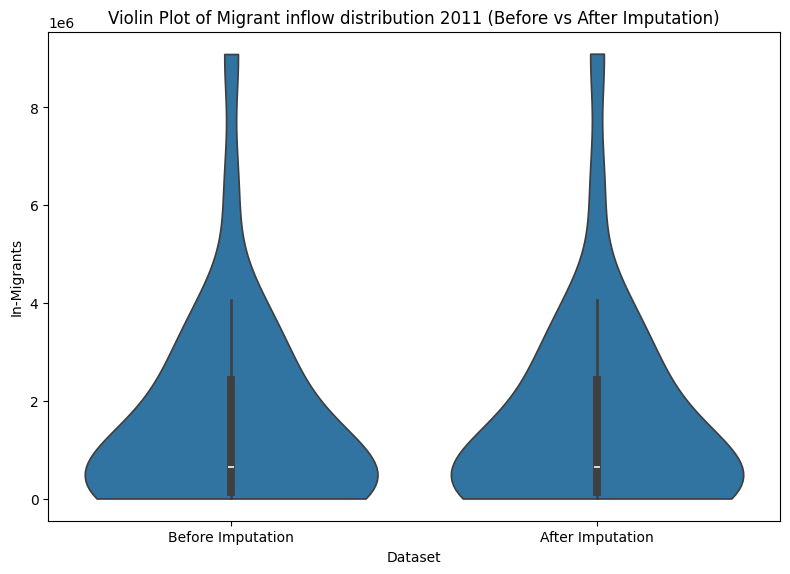}}
  \subfloat[Outflow comparison\label{1b}]{%
        \includegraphics[width=0.37\linewidth]{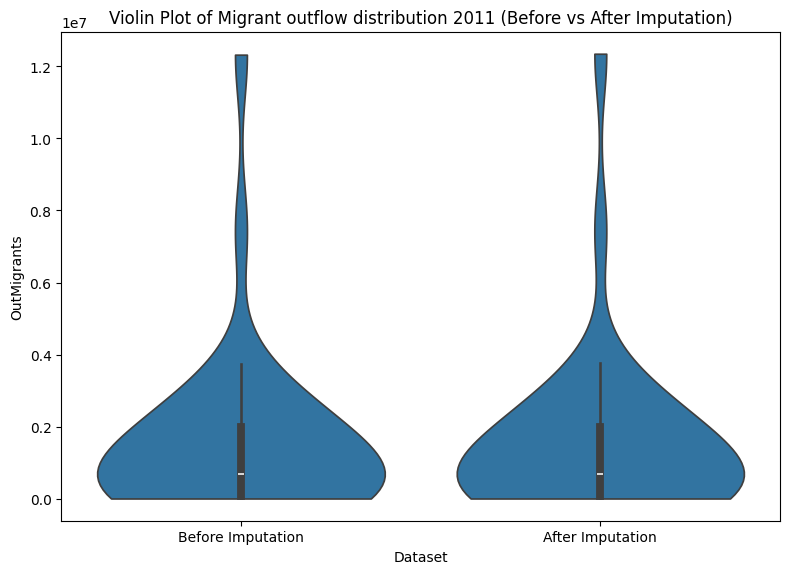}}
    \hfill
  \caption{Violin plots for comparison of 2011 data before and after redistributing ``unclassifiable'' migrants.}
  \label{violin_first_quartile} 
\end{figure}

Table \ref{tab:temporal_smoothness_step1} provides a comparison of the data before and after imputation for the 2011 dataset to demonstrate the impact of this redistribution. It presents a two-step comparison, one for the in-migrant population, based on the place of enumeration, and the other for the place of last residence, considered as the out-migrant population. The two columns show the overall statistical summary of both the in-migrant and out-migrant data distributions before and after imputation. The first three quartile ranges (Q1, Q2, and Q3) across both in-migrant and out-migrant distributions are preserved after imputation. The deviation is observed only in the fourth quartile range, hence drifting the mean and standard deviation. Figure \ref{violin_first_quartile} visualizes the overall data distribution in both cases (in-migrant and out-migrant). The sub-figures display identical violin plots before and after imputation, indicating that the performed redistribution provides completeness in the datasets while preserving the overall data distributions.


\begin{table}[b]
\centering
\caption{Evolution of Interstate Migration networks across decades}
\label{tab:unclassifiable_application}
\begin{tabular}{l|ccc}
\textbf{Property} & \textbf{1991} &\textbf{2001} & \textbf{2011}\\ \hline
Number of nodes&32&35&35\\
Number of edges&992&1190&1190\\
Avg edge weight&26.46$\times 10^{3}$&33.60$\times 10^{3}$&44.61$\times 10^{3}$\\
No. of communities&3&4&4\\
Highest in-migration&Maharashtra&Maharashtra&Maharashtra\\
Highest out-migration&Uttar Pradesh&Uttar Pradesh&Uttar Pradesh\\

\end{tabular}
\end{table}

\subsection{Application of Harmonized datasets}
\begin{figure}[t] 
    \centering
  \subfloat[2011\label{1a}]{%
       \includegraphics[width=0.28\linewidth]{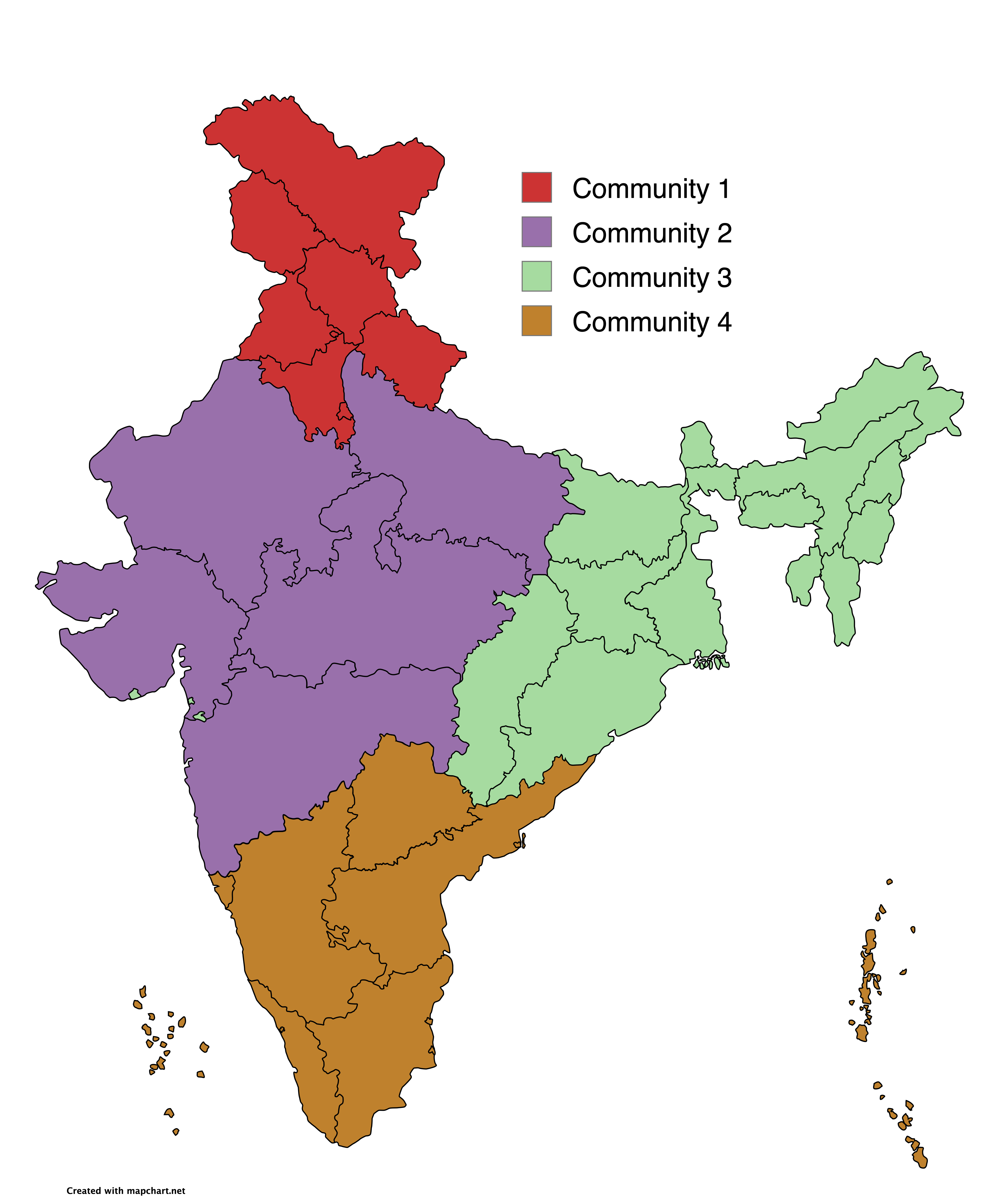}}
  \subfloat[2001\label{1b}]{%
        \includegraphics[width=0.28\linewidth]{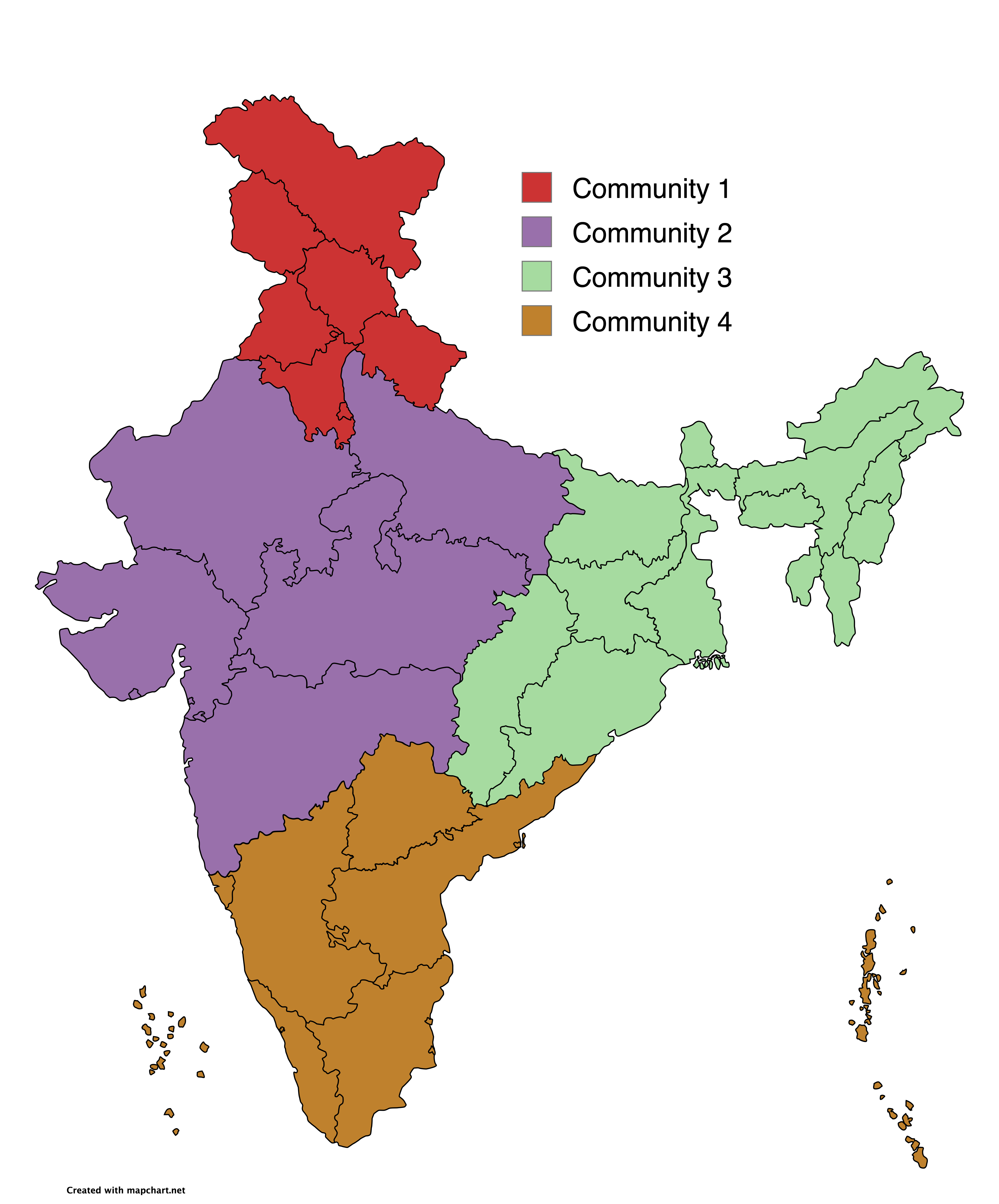}}
  \subfloat[1911\label{1c}]{%
        \includegraphics[width=0.28\linewidth]{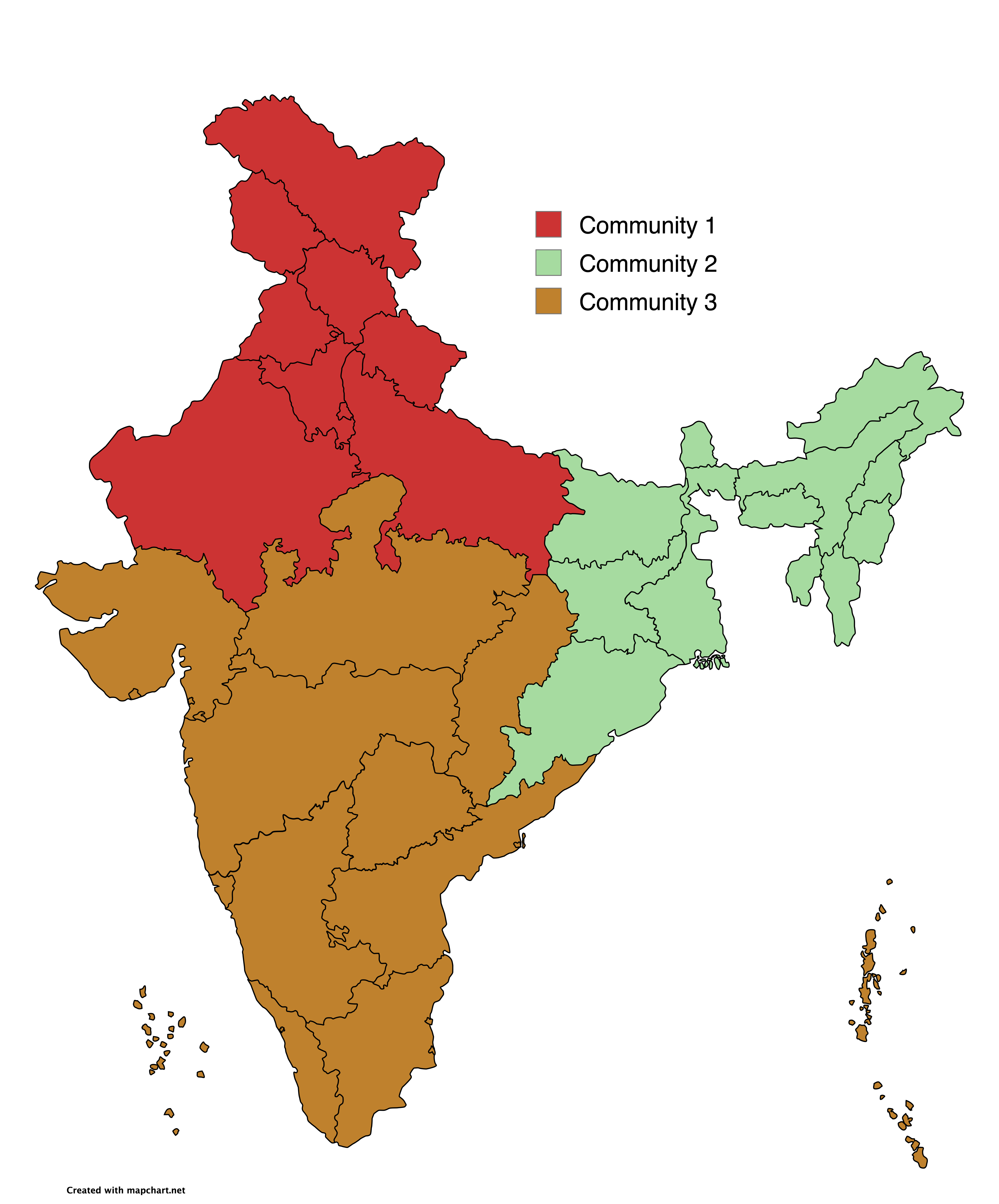}}
    \hfill
  \caption{Communities of migrants evolving over three decades (based on total interstate migrant stocks).}
  \label{communities} 
\end{figure}

An example application of the imputed datasets has been provided by creating Indian interstate migration networks across all three decades. Table \ref{tab:unclassifiable_application} summarizes the properties of these migration networks evolving over three decades. The average edge weight increases by $\approx 30\%$ each decade, indicating an increase in interstate migration over decades. Maharashtra remains the dominant state in receiving the majority of migrants across all three decades, and Uttar Pradesh remains the dominant state in terms of providing out-migrants across all three decades. The interstate migrant communities are identified across all three decades using the Louvain community detection algorithm to identify groups of states with a higher exchange of migrants within the group. As presented in Fig. \ref{communities}, the imputed 1991 dataset reveals three communities of states and union territories. However, the imputed 2001 and 2011 datasets indicate four communities, and both remain stable across these two decades.

\begin{figure}[t] 
    \centering
  \subfloat[1991 Non-Imputed\label{2a}]{%
       \includegraphics[width=0.27\linewidth]{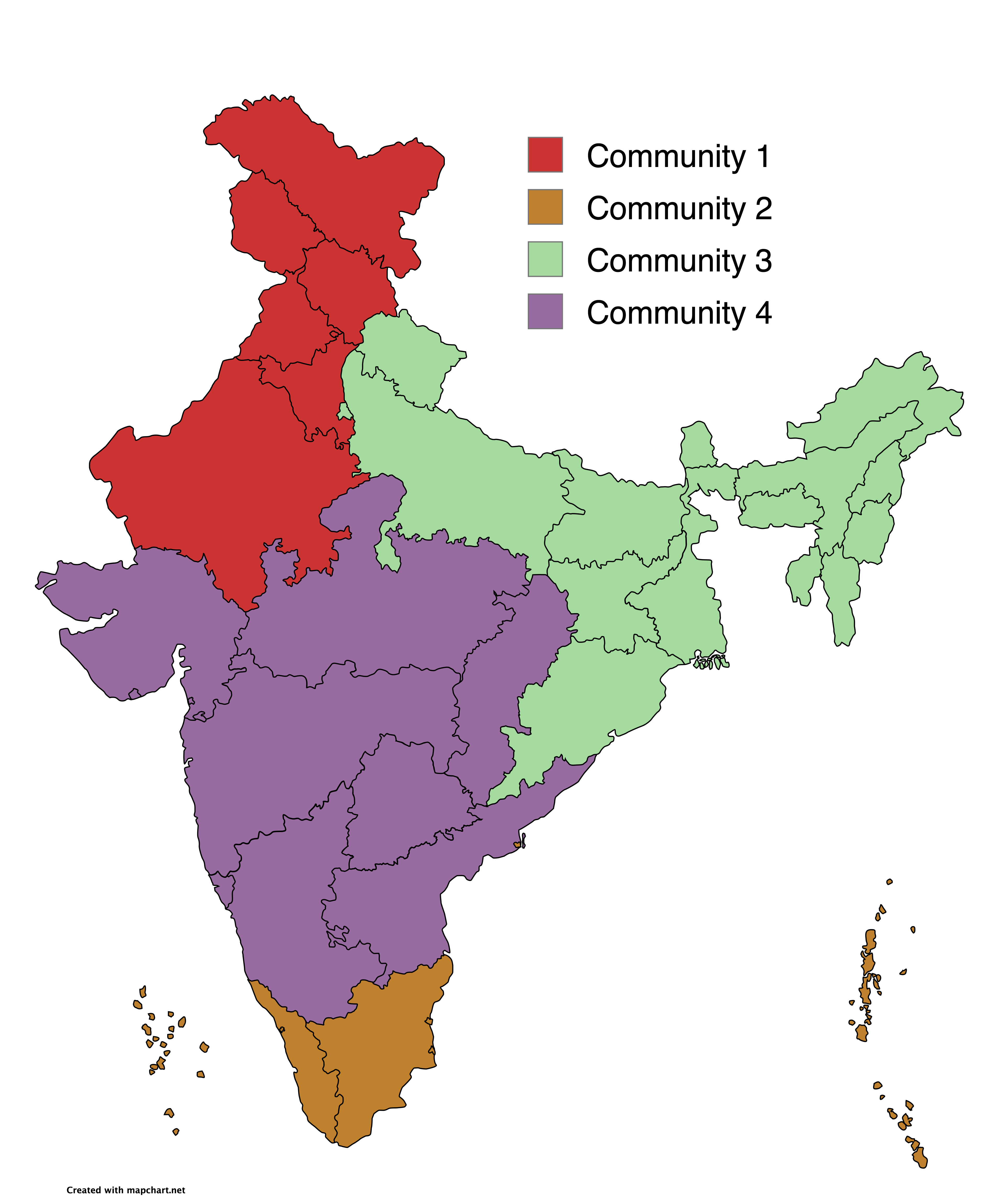}}
  \subfloat[1991 Imputed\label{2b}]{%
        \includegraphics[width=0.27\linewidth]{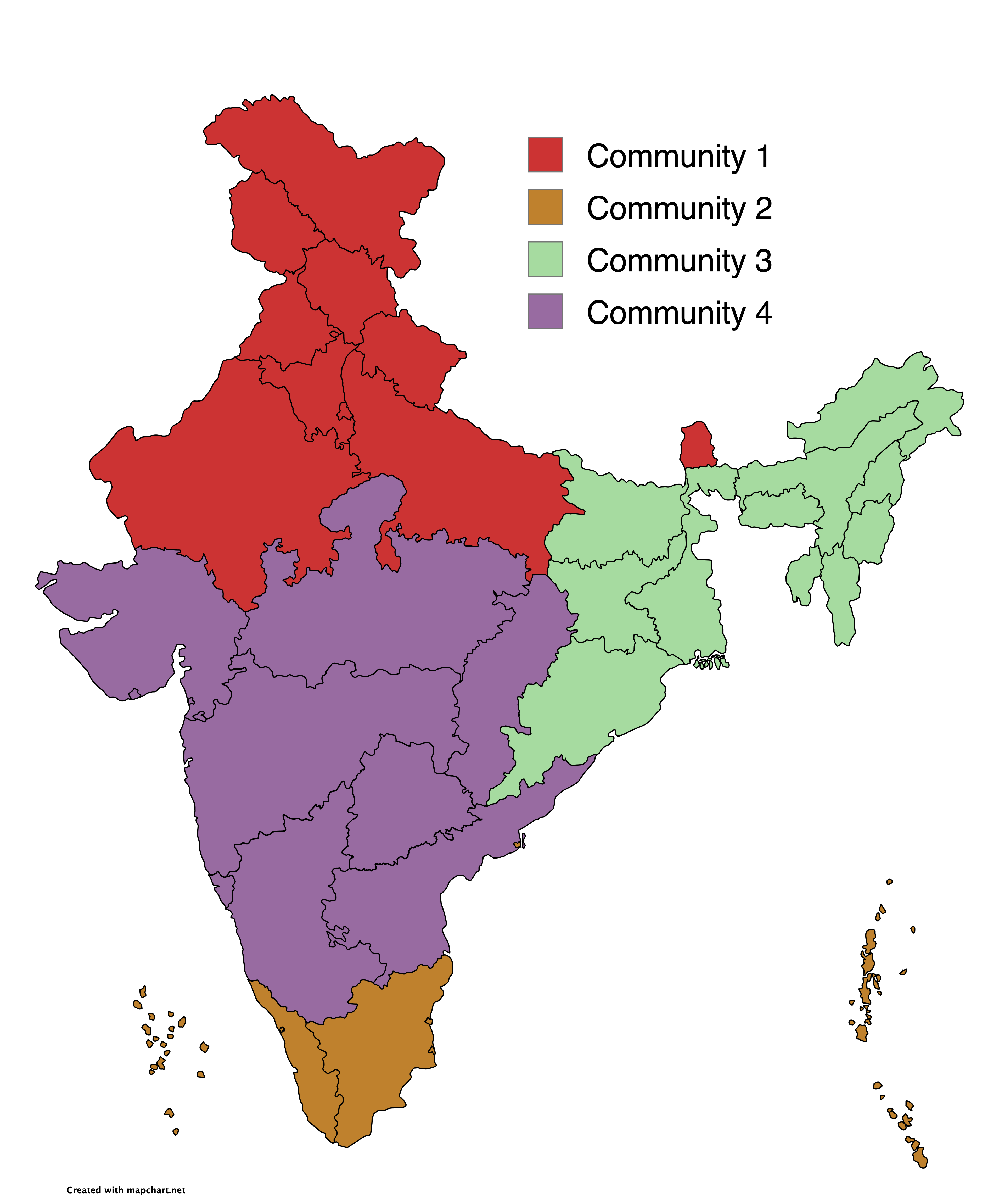}}
    \hfill
  \caption{Communities of migrants in the 1991 dataset before and after imputation}
  \label{communities_Lessthan1year} 
\end{figure}

We perform the same process over the data specific to migrants residing in the state of enumeration for less than one year to observe the community structures of short-duration migrants. Fig. \ref{communities_Lessthan1year} showcases the impact of imputation on the community detection results. In both cases, the number of communities is four, indicating the changing interstate migration trends in one year, deviating from the trends observed from aggregate migration stocks in Fig. \ref{1c}. This application provides a temporal exploratory analysis of the interstate migration networks, ensuring robustness and comparability across decades.

\section{\textbf{Discussion}} \label{discussion}

Our proposed solution systematically addresses the gaps in census-based migration data for Indian interstate migration, utilizing mathematically transparent and constraint-preserving redistribution strategies. This correction enhances the reliability of longitudinal studies, such as state-level migration analyses, facilitating clearer differentiation between source-dominated, destination-dominated, and intermediary states within the migration network. This also improves the validity of network-derived measures such as centrality, flow asymmetry, and community structure, which are sensitive to missing/ misallocated stocks and flows.

The discussed implementation results are focused on the total interstate migrant population (i.e., where the place of enumeration, as well as the place of last residence, is either a state or a union territory) to demonstrate the application of harmonized data. The interstate migration data is extracted as a subset of the entire Indian migration data. The census also reports migration data at other demographic levels, such as by sex, rural/urban residence, intrastate migration, and migration outside India. Thus, the applications of harmonized data can be extended to address analyses providing finer-grained demographic stratification.

\section{\textbf{Conclusion and Future Work}} \label{conclusion}

This paper proposes a reproducible and data-centric framework to identify and correct structural biases in Indian interstate migration data from the Census of India. By addressing migration-specific issues, such as missing source states, temporally inconsistent last places of residence and places of enumeration, and gaps in duration reporting, the framework produces harmonized datasets that preserve census totals while allowing consistent longitudinal and network-based analyses. The resulting imputed datasets can be used by migration researchers, network scientists, urban planners, and policymakers to study interstate migration patterns and regional inequalities. These datasets support long-term migration analysis and the development of policy-aware, fairness-sensitive migration models, such as using multiplex migration networks. As future censuses become increasingly complex and policy-relevant, data-centric, reproducible correction frameworks like the one proposed here will be essential for maintaining longitudinal consistency, ensuring analytical reliability, and promoting responsible use.



\bibliographystyle{unsrt}  
\bibliography{references}

@article{gupta2020study,
  title={Study of Changing Trends and Patterns of Internal Migration and Factors Affecting It},
  author={Gupta, Anushka and Miglani, Mehak},
  journal={International Journal},
  volume={1},
  number={4},
  year={2020}
}

@book{kumar2024numbers,
  title={Numbers as political allies: The census in Jammu and Kashmir},
  author={Kumar, Vikas},
  year={2024},
  publisher={Cambridge University Press}
}

@incollection{batra2024computational,
  title={A Computational Study of Indian Interstate Migration through the Gender Lens},
  author={Batra, Niveditta and Chaudhuri, Mayurakshi and Chattopadhyay, Chiranjoy},
  booktitle={India Migration Report 2023},
  pages={314--335},
  year={2024},
  publisher={Routledge India}
}

@article{islam2024enquiry,
  title={An Enquiry into the Nature and Problems of Migration in India: A Critical Look},
  author={Islam, Mahibul and Saha, Subrata},
  journal={International Research Journal of Economics and Management Studies IRJEMS},
  volume={3},
  number={3},
  year={2024}
}

@incollection{borhade2025internal,
  title={Internal labour migration in India: Emerging needs of comprehensive national migration policy},
  author={Borhade, Anjali},
  booktitle={Internal migration in contemporary India},
  pages={212--252},
  year={2025},
  publisher={Routledge India}
}

@incollection{bhagat2025nature,
  title={Nature of Migration and its contribution to India's urbanisation},
  author={Bhagat, RB},
  booktitle={Internal Migration in Contemporary India},
  pages={23--40},
  year={2025},
  publisher={Routledge India}
}

@article{datta2013indian,
  title={Indian census data on migration},
  author={Datta, Pranati},
  journal={The Oriental Anthropologist},
  volume={13},
  number={1},
  pages={17--22},
  year={2013},
  publisher={SAGE Publications Sage India: New Delhi, India}
}

@article{garha2019indian,
  title={Indian diaspora population and space: national register, UN Global Migration Database and Big Data},
  author={Garha, Nachatter Singh and Domingo, Andreu},
  journal={Diaspora Studies},
  volume={12},
  number={2},
  pages={134--159},
  year={2019},
  publisher={Brill}
}

@incollection{kumar2023census,
  title={When census is an election: A game-theoretic analysis of over-reporting of headcount},
  author={Kumar, Vikas},
  booktitle={Power and Responsibility: Interdisciplinary Perspectives for the 21st Century in Honor of Manfred J. Holler},
  pages={373--393},
  year={2023},
  publisher={Springer}
}

@article{kumar2020census,
  title={Census laws and the quality of census data: The limits of punitive legislation},
  author={Kumar, Vikas},
  journal={Statistical Journal of the IAOS},
  volume={36},
  number={4},
  pages={1143--1160},
  year={2020},
  publisher={SAGE Publications Sage UK: London, England}
}

@article{yadav2020quality,
  title={The quality of age data: Comparison between two recent Indian censuses 2001--2011},
  author={Yadav, Akhilesh and Vishwakarma, Minakshi and Chauhan, Shekhar},
  journal={Clinical Epidemiology and Global Health},
  volume={8},
  number={2},
  pages={371--376},
  year={2020},
  publisher={Elsevier}
}

@article{starr2021measuring,
  title={{Measuring caste in India}},
  author={Starr, Kelsey Jo and Sahgal, Neha},
  journal={Pew Research Center (Decoded), June},
  volume={29},
  year={2021}
}

@article{das2015nativity,
  title={Is nativity on rise? Estimation of interstate migration based on census of india 2011 for major States in India},
  author={Das, Bhaswati and Mistri, Avijit},
  journal={Social Change},
  volume={45},
  number={1},
  pages={137--144},
  year={2015},
  publisher={SAGE Publications Sage India: New Delhi, India}
}

@article{tumbe2012migration,
  title={Migration persistence across twentieth century India},
  author={Tumbe, Chinmay},
  journal={Migration and Development},
  volume={1},
  number={1},
  pages={87--112},
  year={2012},
  publisher={SAGE Publications Sage India: New Delhi, India}
}

@techreport{sahai2023social,
  title={Social Networks and Internal Migration: Evidence from Facebook in India},
  author={Sahai, Harshil and Bailey, Michael},
  year={2023},
  institution={Working Paper}
}

@article{datta2024spatial,
  title={Spatial Patterns of Heaping in Age Data among Literates, Illiterates, and Numeracy--Literacy Correlates: A Cross-Sectional Analysis of Census 2011, of India},
  author={Datta, Jayanta and Sinha, Prasenjit},
  journal={Indian Journal of Community Medicine},
  volume={49},
  number={1},
  pages={189--194},
  year={2024},
  publisher={Medknow}
}

@article{singh2025counting,
  title={Counting Power: The Caste Census and the Four Faces of Exclusion in India},
  author={Singh, Yasha and Ranjan, Nitin},
  year={2025},
  publisher={Inside Jharkhand}
}

@article{singh2025caste,
  title={Caste Census Data for a Just Republic},
  author={Singh, Yasha and Ranjan, Nitin},
  year={2025}
}

@article{ranjan2025caste,
  title={Caste Census: A Detailed Report},
  author={Ranjan, Nitin and Singh, Yasha},
  year={2025}
}

@article{sircar2017census,
  title={‘Census Towns’ in India and what it means to be ‘urban’: Competing epistemologies and potential new approaches},
  author={Sircar, Srilata},
  journal={Singapore Journal of Tropical Geography},
  volume={38},
  number={2},
  pages={229--244},
  year={2017},
  publisher={Wiley Online Library}
}

@incollection{chattopadhyay2024age,
  title={Age at Marriage of Indian Women},
  author={Chattopadhyay, Aparajita and Singh, Akancha},
  booktitle={Atlas of Gender and Health Inequalities in India},
  pages={115--123},
  year={2024},
  publisher={Springer}
}

@article{srivastava2011internal,
  title={Internal migration in India},
  author={Srivastava, Ravi},
  journal={Human Development in India},
  year={2011},
  publisher={UNESCO}
}

@article{srivastava2013impact,
  title={Impact of internal migration in India},
  author={Srivastava, Ravi},
  journal={Refugee and Migratory Movements Research Unit (RMMRU). http://www. rmmru. org/newsite/wp-content/uploads/2013/08/workingpaper41. pdf Social Security for Internal Migrants},
  volume={237},
  year={2013}
}

@article{coffey2015short,
  title={Short-term labor migration from rural north India: Evidence from new survey data},
  author={Coffey, Diane and Papp, John and Spears, Dean},
  journal={Population Research and Policy Review},
  volume={34},
  number={3},
  pages={361--380},
  year={2015},
  publisher={Springer}
}

@article{kundu2019trends,
  title={Trends in mobility in India: issues of labour market integration and exclusion of vulnerable sections of the population},
  author={Kundu, Amitabh},
  journal={Area Development and Policy},
  volume={4},
  number={4},
  pages={346--366},
  year={2019},
  publisher={Taylor \& Francis}
}

@incollection{korra2012short,
  title={Short-duration migration in India},
  author={Korra, Vijay},
  booktitle={India Migration Report 2011},
  pages={52--71},
  year={2012},
  publisher={Routledge India}
}

\end{document}